# Topological nodal-line fermions in ZrSiSe and ZrSiTe

Jin Hu[1][†], Zhijie Tang[1][†], Jinyu Liu[1], Xue Liu[1], Yanglin Zhu[1], David Graf[2], Yanmeng Shi[3], Shi Che[3], Chun Ning Lau[3], Jiang Wei[1] and Zhiqiang Mao[1]*

[1]Department of Physics and Engineering Physics, Tulane University, New Orleans, LA 70118, USA

[2]National High Magnetic Field Lab, Tallahassee, FL 32310, USA

[3]Department of Physics, University of California, Riverside, CA 92521, USA

**Abstract**

**The discovery of topological semimetal phase in three-dimensional (3D) systems is a new breakthrough in topological material research. Dirac nodal-line semimetal is one of the three topological semimetal phases discovered so far; it is characterized by linear band crossing along a line/loop, contrasted with the linear band crossing at discrete momentum points in 3D Dirac and Weyl semimetals. The study of nodal-line semimetal is still at initial stage; only three material systems have been verified to host nodal line fermions until now, including PbTaSe$_2$ [1], PtSn$_4$ [2] and ZrSiS [3]. In this letter, we report evidence of nodal line fermions in ZrSiSe and ZrSiTe probed in de Haas–van Alphen (dHvA) quantum oscillations. Although ZrSiSe and ZrSiTe share similar layered structure with ZrSiS, our measurements of angular dependences of dHvA oscillations indicate the Fermi surface (FS) enclosing Dirac nodal line is of 2D character in ZiSiTe, in contrast with 3D-like FS in ZrSiSe and ZrSiS. Another important property revealed in our experiment is that the nodal line fermion density in ZrSi(S/Se) (~ $10^{20}$-$10^{21}$ cm$^{-3}$) is much higher than the Dirac/Weyl fermion density of any known topological materials. In addition, we have demonstrated ZrSiSe and ZrSiTe single crystals can be thinned down to 2D atomic thin**

**layers through microexfoliation, which offers a promising platform to verify the predicted 2D topological insulator in the monolayer materials with ZrSiS-type structure** [4].

Topological semimetal phases discovered in three-dimensional (3D) materials represent new quantum states of matter, which have attracted intensive studies in recent years. There have been several different forms of topological semimetals established experimentally, including Dirac semimetals, Weyl semimetals and topological nodal-line semimetals. These materials exhibit technologically useful properties such as high bulk carrier mobility and large magnetoresistance. In 3D Dirac semimetals such as $Na_3Bi$ [5-6] and $Cd_3As_2$ [7-11], four-fold degenerate linear band crossings at Dirac nodes occur at discrete momentum points and are protected against gap opening by crystalline symmetry. Dirac semimetals can be regarded as parent materials of Weyl semimetals. With lifted spin degeneracy by either broken time reversal or broken spatial inversion symmetry, each Dirac cone splits into a pair of Weyl cones with opposite chirality [5, 7, 12-13]. Weyl states have been demonstrated in many materials, including (Ta/Nb)(As/P) [14-19], (Mo/W)$Te_2$ [20-30], LaAlGe [31] and YbMn$Bi_2$ [32]. For topological nodal-line semimetals, Dirac bands crossing takes place along a one-dimensional line in momentum space, contrasted with Dirac band crossing at discrete momentum points in Dirac or Weyl semimetals. The experimentally established examples of topological nodal-line semimetal include PbTa$Se_2$ [1], PtS$n_4$ [2] and ZrSiS [3, 33].

Among the reported topological nodal-line semimetals, ZrSiS was found to have other distinct properties. First, it also harbors a new type of theoretically predicted Dirac cone state which is generated by a square lattice and protected by the non-symmorphic symmetry in addition to the nodal-line Dirac cone [34]. Second, its Dirac bands are linearly dispersed in a wide energy range, up to 2eV, much larger than that of any other known Dirac materials [3, 33]. Third, unusual surface states hybridized with bulk bands have also been probed in ZrSiS [3]. Moreover, surprisingly strong Zeeman splitting effect has been probed in the de Haas–van Alphen (dHvA) effect [35]. These fascinating properties motivated us to investigate ZrSiSe and ZrSiTe, which are isostructural to ZrSiS. These three compounds belong to a large family of materials WHM with the PbFCl-type structure (W= Zr, Hf, or La; H=Si, Ge, Sn or Sb; M=O, S, Se and Te). The first-principle calculations by Xu *et al* [4] have shown that these isostructural compounds indeed host similar electronic structures. One distinct aspect of this group of materials is that the two-dimensional (2D) topological insulator (TI), which has been proved only in a few systems such as HgTe/CdTe [36-37] and InAs/GaSb [38] quantum wells, can possibly be realized in monolayer films of WHM; ZrSiO monolayer is the first predicted example [4]. Recent angle-resolved photoemission (ARPES) studies on ZrSnTe, which is also a member of WHM, by Lou *et al*. [39] has revealed evidence of TI band structure on the top layer of this material. When such a predicted 2D TI is realized experimentally, it would offer a wonderful platform for testing a new device concept based on 2D TI – "topotronic" device [4]. Another important result predicted by Xu *et al.* [4] is that the interlayer bonding energy, as well as particle-hole asymmetry, varies strikingly with the M atom. The interlayer binding energy for ZrSiO is 1.1664 eV, but decreases to 0.2131 eV and 0.0593 eV for ZrSiSe and ZrSiTe respectively, implying the structural dimensionality evolution toward 2D, whereas the particle-hole asymmetry increases accordingly [4]. The low interlayer binding energy can possibly make atomically thin layers of these two materials accessible through mechanical exfoliation. If this proves to be true, these materials would provide a rare opportunity to investigate 2D TIs.

In this letter, we report the discovery of nodal-line fermions in ZrSiSe and ZrSiTe, and demonstrated the accessibility of their atomically thin layers. Through the observation and analyses of dHvA quantum oscillations, we have found evidences of Dirac fermions from multiple Dirac bands for both materials, including non-trivial Berry phase of cyclotron orbit, high quantum mobility and light effective mass. We have also examined the Fermi surface (FS) dimensionality of ZrSiSe and ZeSiTe by measuring the angular dependences of dHvA and Shubnikov-de Hass (SdH) oscillation frequencies. We found that ZrSiSe shares a similar 3D-like FS with ZrSiS, whereas the FS of ZrSiTe shows a more remarkable 2D-like signature though a 3D-like component is also observed. Another important result revealed in our experiments is the high density of Dirac nodal-line fermions in ZrSiM ($\sim 10^{20}$-$10^{21}$ cm$^{-3}$ for M=S and Se), 2-3 orders of magnitude higher than those of other known Dirac materials with discrete Dirac nodes. Our findings not only pave a way to further understand novel exotic properties of nodal-line semimetal phase in the WHM family, but also highlight a new possible route to realize 2D TIs.

The plate-like ZrSiSe and ZrSiTe single crystals (Fig. 1c) were synthesized using a chemical vapor transport method similar to that used for ZrSiS single crystal growth [35]. The excellent crystallinity of these crystals is demonstrated by the sharp (00L) reflections in the X-ray diffraction spectra shown in Fig. 1b. As shown in Fig. 1a, all ZrSiM (M=S, Se, Te) compounds share a similar tetragonal structure formed from the stacking of M-Zr-Si-Zr-M slabs [40]; the slab thickness along the *c*-axis extracted from the X-ray diffraction measurements (Fig. 1b) is increased by 4% and 18% in ZrSiSe and ZrSiTe respectively, as compared to ZrSiS. This is caused by the increased ionic radius from S to Te ions, which adjusts a delicate steric-electronic balance and results in the structural dimensionality evolution from 3D to 2D [4, 40] as mentioned above.

In spite of reduced dimensionality, ZrSiSe and ZrSiTe also host nodal-line fermions, which has been evidenced by our quantum oscillation experiments. In Figs. 1d-1f, we present the isothermal magnetization measured up to 7T for ZrSiSe (Figs. 1d-1e) and ZrSiTe (Fig. 1f) single crystals. In ZrSiSe, both the out-of-plane ($B//c$, Fig. 1d) and in-plane ($B//ab$, Fig. 1e) magnetizations exhibit strong dHvA oscillations, which extend to $B$=2T and 1T for $B//c$ and $B//ab$ respectively and sustain up to $T$ =20K. Similar features were also observed in Dirac nodal-line semimetal ZrSiS whose FS consisting of the nodal line has been shown to be of 3D character. This implies that the FS of ZrSiSe is also of a 3D character [35]. In ZrSiTe, however, although quantum oscillations are present in the out-of-plane magnetization (Fig. 1f), it is hardly seen when the field is oriented along the in-plane direction, suggesting that ZrSiTe may have 2D-like FS. More detailed discussions on the evolution of Fermi surface from ZrSiS to ZrSiSe and ZrSiTe will be given below.

From the analyses of dHvA oscillations, we have seen typical signatures of Dirac fermions in both ZrSiSe and ZrSiTe. We first focus on ZrSiSe whose lattice structure is only slightly different from that of ZrSiS (Fig. 1a). In Figs. 2a and 2d, we present the oscillatory components of magnetization $\Delta M$ for ZrSiSe, obtained after subtracting the background. Strong oscillations with the amplitudes of 1~2 emu/mol at $T$=2K and $B$=7T are clearly seen for both out-of-plane ($B//c$, Fig. 2a) and in-plane ($B//ab$, Fig. 2d) field orientations. From the Fast Fourier transformation (FFT) analyses, we have derived a single frequency of 210T for out-of-plane oscillations (Fig. 2b, inset), which is comparable with the dHvA oscillation frequency $F_\beta$ (=240T) probed in ZrSiS for the same field orientation [35]. Nevertheless, in addition to $F_\beta$, ZrSiS also shows a very small frequency ($F_\alpha$=8.4T) in dHvA oscillations for $B//c$, which was not observed in ZrSiSe. This implies that the FSs of ZrSiS and ZrSiSe have somewhat different morphologies though both share a common 3D-like character. In sharp contrast with the single frequency oscillations for $B//c$, the in-plane dHvA oscillation pattern (Fig. 2d) consists of the superposition of both high- and low-frequency oscillation components. As shown in Fig. 2f, we have separated the low (upper panel) and high (lower panel) frequency components. Both components display beat patterns, indicating that each component consists of multiple frequencies. Indeed, FFT analyses (Fig. 2e, inset) have revealed three major low frequencies (19.2T, 22.8T, and 24T) and three major large frequencies (126.9T, 132.7T, and 142T). Multiple frequencies under in-plane field have also been observed in ZrSiS [35], which is an indication of electronic structure with 3D character.

In general, the oscillatory magnetization of a Dirac system can be described by Lifshitz-Kosevich (LK) formula [41] which takes Berry phase into account [42]:

$$\Delta M \propto -B^{\lambda} R_T R_D R_S \sin[2\pi(\frac{F}{B}+\gamma-\delta)] \quad (1)$$

where $R_T = \alpha T m^*/B\sinh(\alpha T m^*/B)$, $R_D = \exp(-\alpha T_D m^*/B)$, and $R_S = \cos(\pi g m^*/m_0)$. $m^*$ and $m_0$ are effective cyclotron mass and free electron mass, respectively. $T_D$ is Dingle temperature, and $\alpha = (2\pi^2 k_B m_0)/(\hbar e)$. The oscillation of $\Delta M$ is described by the sine term with a phase factor $\gamma - \delta$, in which $\gamma = \frac{1}{2} - \frac{\phi_B}{2\pi}$ and $\phi_B$ is Berry phase. The phase shift $\delta$ is determined by the dimensionality of Fermi surface; $\delta = 0$ and $\pm 1/8$, respectively, for 2D and 3D cases. In addition, the exponent $\lambda$ is also determined by the dimensionality; $\lambda = 1/2$ and 0 for 3D and 2D cases, respectively. From the LK formula, the effective electron cyclotron mass $m^*$ can be obtained through the fit of the temperature dependence of the oscillation amplitude to the thermal damping factor $R_T$, as shown in Figs. 2b and 2e. For all probed oscillation frequencies, the obtained effective mass are in the range of 0.04-0.08 $m_0$ (see Table 1), which are comparable to that of ZrSiS [35] and Dirac semimetal $Cd_3As_2$ [43-44]. Using the fitted effective mass and the oscillation frequencies derived from the FFT spectra as known parameters, we can further fit the oscillation patterns at a given temperature (*e.g.* $T$=1.8K) to the LK formula (Eq. 1), from which quantum mobility and Berry phase can be extracted. We have adopted 3D LK formula given that the FS of ZrSiSe is found to exhibit strong 3D character (see below). As shown in Fig. 2c, the 3D LK formula describes the single-frequency out-of-plane dHvA oscillations very well, yielding the Dingle temperature $T_D$ of 5.6K and the phase factor $\gamma - \delta$ of 0.53. The quantum relaxation time $\tau_q$ [= $\hbar/(2\pi k_B T_D)$] corresponding to $T_D$ = 5.6K is 2.2 ×$10^{-13}$ s, from which the quantum mobility $\mu_q$ [= $e\tau/m^*$] is estimated to be 4605 $cm^2/(Vs)$. The Berry phase $\phi_B$ determined from the phase factor is -2π(-0.03+$\delta$). As we show below, the FS associated with the $F$ = 210T oscillations possesses a dimensionality between 2D and 3D, implying that the value of $\delta$ should be between 0 and ±1/8. This places $\phi_B$ in between -0.31π and -0.19π, indicating a non-trivial Berry phase. Similar results can also be obtained through the more commonly used Landau level (LL) fan diagram (Supplementary Fig. 1). However, it is worth noting that the results obtained from the direct LK-fit should be more reliable, since we can only achieve a LL index of 32 for the highest field of 7T, which leads the intercept obtained from the extrapolation of the linear fit in the LL fan diagram to have a larger uncertainty.

For the in-plane oscillations which contains several frequencies, the fits have to be done using the multiband LK formula for which the total oscillations with multiple frequencies can be treated as linear superposition of the oscillations with various frequencies. Such an approach has been shown to be effective for analyzing the dHvA oscillations of ZrSiS [35] and SdH oscillations of Weyl semimetal TaP [45]. As stated above, the in-plane dHvA oscillations of

ZrSiSe consist of low- and high-oscillation components; both are characterized by multiple frequencies. In order to achieve more accurate fits, we have separated the low- and high-frequency components and fit them individually, as shown in Fig. 2f. These fits yield the Dingle temperature of 5-15K, as shown in Table 1. Combined with the effective masses obtained in Fig. 2e, we have derived quantum mobility ranging from 2500 cm$^2$/Vs to 9500 cm$^2$/Vs for various frequencies, as listed in Table 1. Additionally, from these multiple-band LK fits, we also obtained non-trivial Berry phase for each band when the dimensionality factor $\delta$ takes either the value of zero for a 2D case or $\pm 1/8$ for a 3D case. These results indicate that electrons involved in the in-plane quantum oscillations are also Dirac-like, similar to the observations in ZrSiS [35].

To make a more comprehensive comparison between ZrSiS and ZrSiSe, we have systematically measured the variation of the oscillation frequencies of ZrSiSe with the magnetic field orientation (see the inset of Fig. 4d for the experiment setup). From these measurements, we can obtain key information of FS morphology, given that the quantum oscillation frequency is directly linked to the extremal FS cross-section area $A_F$. As presented in Fig. 4a, after the background subtraction, the oscillation patterns display a clear evolution with the rotation of magnetic field. From the FFT analyses (Supplementary Fig. 2a), we have determined the angular dependences of the oscillation frequencies, as shown in Fig. 4b. Overall, we have observed three major frequency branches. With increasing the angle $\theta$, the high frequency branch ($F_\beta$), which corresponds to the 210T single frequency for $B//c$ (Fig. 2b), continues to increase until it disappears above 68°. Such an angular dependence can be fitted with $F_{2D}/\cos\theta + F_{3D}$ which includes both 2D and isotropic 3D FS components. The relative weight between the 2D and 3D components derived from the fit, $F_{2D}/F_{3D}$, is found to be ~ 0.8, indicating that the dimensionality of the FS associated with $F_\beta$ is between 2D and 3D. In addition to $F_\beta$, another two low frequency branches with weak angular dependences ($F_\eta$ and $F_\alpha$) are also present for 45° < $\theta$ < 90° (Fig. 4b), indicating a 3D component of the FS. Such angular dependences of $F_\beta$, $F_\eta$ and $F_\alpha$ in ZrSiSe have also been verified by SdH effects. The SdH oscillations, though much weaker than dHvA oscillations, can be observed in high field magnetotransport experiments. Fig. 4e shows the magnetoresistance of ZrSiSe as a function of magnetic field measured up to 31T under various field orientations; SdH oscillation patterns can be resolved after subtracting the background (Supplementary Fig. 3a). Oscillation frequencies extracted from the FFT spectra of SdH oscillations (Supplementary Fig. 3b) are in good agreement with the results measured by the dHvA oscillations, as shown in Fig.4b. Moreover, we note ZrSiSe exhibits large magnetoresistance, with a "butterfly-shaped" anisotropy (Supplementary Fig. 5). The maximum value reaches ~4,000% at 31T and $\theta$ ~33°. Similar feature has also been reported in ZrSiS [46-49].

In our previous dHvA studies on ZrSiS [35], we observed angular dependences of frequencies similar to $F_\beta$ and $F_\eta$ seen in ZrSiSe. However, although ZrSiS also exhibits a low frequency branch $F_\alpha(\theta)$ (~8-20T), it displays features distinct from that seen in ZrSiSe. $F_\alpha$ of ZrSiS shows a continuous evolution from $\theta$ = 0° (B//c) to $\theta$ = 90 ° (B//ab); it can be fitted to $F_\alpha(\theta=0)/\cos\theta$ in low angle region (0° < $\theta$ < 62°), but becomes weakly angle dependent for 62° < $\theta$ < 90°, suggesting that the FS associated with $F_\alpha$ in ZrSiS is of a striking 2D character though a 3D component also exists [35]. In contrast, $F_\alpha(\theta)$ of ZrSiSe is present only in the 45° < $\theta$ < 90° angle range and remains nearly constant (Fig. 4b), indicating the $F_\alpha$ FS of ZrSiSe has distinct morphology from that of ZrSiS. APRES studies have shown that the FS of ZrSiS consists of four lens-shaped hole pockets surrounding the Γ point and small electron pockets at X point [3, 33]. The Dirac nodal-lines are enclosed by the hole pockets, while the electron pocket at X involves hybridization of surface states and the bulk bands associated with the Dirac cone protected by non-symmorphic symmetry [3]. The high oscillation frequency $F_\beta$ in ZrSiS is found to arise from the lens-shaped FS pockets, while the low frequency $F_\alpha$ corresponds to the small electron pocket at X [35]. Given similar crystal (Fig. 1a) and electronic structures between ZrSiS and ZrSiSe [4], we can reasonably expect existence of a similar lens-shaped FS enclosing Dirac nodal-line in ZrSiSe. The $F_\beta$ and $F_\eta$ branches probed in our dHvA oscillations of ZrSiSe (Fig. 4b) most likely correspond to the expected nodal-line FS. On the other hand, although the major features of band structures are similar between ZrSiS and ZrSiSe, replacing S with Se should more or less modify the FS. This may explain distinct characteristics of $F_\alpha$ between these two compounds. In addition, compared with ZrSiS, Dirac fermions in ZrSiSe become more massive (see Table 1, $B//c$), which may be ascribed to the fact that the particle-hole asymmetry is enhanced [4] and the spin-orbit coupling (SOC) induced gap at Dirac nodal line/point is larger in ZrSiSe.

Next we discuss our results obtained on ZrSiTe. Although ZrSiSe preserves major quantum oscillation properties of ZrSiS as discussed above, we expect more significant electronic structure modifications in ZrSiTe due to enhanced lattice expansion (Fig. 1b) and reduced interlayer coupling [4]. As indicated above, the interlayer binding energy of ZrSiTe is significantly reduced as compared with ZrSiS [4], which should lead to more 2D-like electronic structure. This expectation is indeed consistent with our observation that dHvA oscillations in ZrSiTe are hardly resolved for in-plane field. Following the approach used for analyzing dHvA effect in ZrSiSe, we have extracted the oscillatory component of the out-of-plane magnetization for ZrSiTe (Fig. 3a), and determined the oscillation frequencies to be 102T and 154T from its FFT spectrum (Fig. 3b, inset). The effective masses corresponding to these two frequencies, obtained from the fit of the temperature dependence of the FFT amplitude to the temperature damping term of the LK formula $R_T$ (Eq. 1), are $0.093m_0$ and $0.091m_0$, respectively (Fig. 3b). The fit of the oscillation pattern to the 3D LK formula (Fig. 3c) yield quantum mobility of 1625 cm$^2$/VS for 102T and 940 cm$^2$/VS for the 154T component as well as non-trivial Berry phase as listed in Table 1. As shown below, ZrSiTe exhibits strong 2D character. We have also fit the oscillation pattern to the 2D LK formula (Eq. 1), and obtained similar phase factor and slightly larger Dingle temperature (Supplementary Fig. 4). These results indicate that electrons participating in quantum oscillations in ZrSiTe are also Dirac fermions.

Given that similar band structures have been predicted for all ZrSiM compounds by first principle calculations [4] and the dHvA oscillation frequency of 154T probed in ZrSiTe is not far from $F_\beta$ probed in ZrSiS ($F_\beta$=240T for $B//c$) and ZrSiSe ($F_\beta$=210T for $B//c$), the FS associated with $F_\beta$, which consists of nodal lines, is most likely preserved in ZrSiTe. As noted above, Dirac fermions in ZrSiTe are more massive as compared with ZrSiS and ZrSiSe. This can probably be associated with the enhanced particle-hole asymmetry [4] and larger SOC-induced gap at the nodal line in ZrSiTe. To examine the FS morphology of ZrSiTe, we have also investigated the evolution of dHvA oscillations with the field rotation and found that the oscillatory component of magnetization is gradually suppressed with increasing $\theta$, becoming hardly resolved for $\theta > 45°$, as shown in Fig. 4c. Fig. 4d gives the evolution of dHvA oscillation frequencies extracted from FFT analyses for $0° \leq \theta \leq 45°$ (see Supplementary Fig. 2b). The higher frequency component exhibits a very weak angular-dependence, while the lower frequency component shows a clear variation with $\theta$. Further, the lower frequency branch can be fitted to $F_{2D}/\cos\theta + F_{3D}$ as illustrated by the red fit curve in Fig. 4d. The relative weight between 2D and 3D components derived from the fit, $F_{2D}/F_{3D}$ is ~ 1.7, about twice as large as that of ZrSiSe ($F_{2D}/F_{3D}$ ~ 0.8), implying that the FS of ZrSiTe is of 2D character though a 3D component also exists. The

2D character of the FS of ZrSiTe is in line with the reduced interlayer binding energy predicted for this material [4, 40].

Additionally, unlike other topological materials characterized by discrete Dirac nodes which leads to low carrier density, ZrSiM compounds are expected to have very high density of Dirac fermions due to the existence of Dirac nodal line, which is crucial for the practical electronics applications. In order to estimate the carrier density of ZrSiSe, we have measured its longitudinal and Hall resistivity. Fig. 4e and 3f present the data. The field dependence of Hall resistivity (Fig. 4f) as well as the temperature dependence of Hall coefficient $R_H$ (Fig. 4g) exhibit clear multiple band signatures. For a multiple band system, we can normally estimate the carrier mobility and density by simultaneously fitting magnetoresistivity and Hall resistivity using a two-band model. However, we did not get a successful fit for ZrSiSe. If we take a single-band approximation, the carrier density $n$ is roughly estimated to be $10^{21}$ cm$^{-3}$. For ZrSiS, we also made a similar estimate using the transport data we reported previously [35] and the estimated $n$ is found to be in the same order of magnitude. Moreover, we also achieved a successful fit using a two-band model for ZrSiS, which gives $n \sim 10^{20}$ cm$^{-3}$, just one order of magnitude less than that estimated using the single band model. The carrier density of $10^{20}$-$10^{21}$ cm$^{-3}$ is significantly higher than those in Dirac semimetals such as Cd$_3$As$_2$ ($10^{18}$ cm$^{-3}$ [43, 50]) and Na$_3$Bi ($10^{17}$ cm$^{-3}$ [51]), topological insulators (typically $10^{10-12}$ cm$^{-3}$ [52]), and graphene with gate voltage applied ($10^{10-12}$ cm$^{-3}$ [53-54]).

The weak interlayer bonding of ZrSiSe and ZrSiTe implies the possibility that atomically thin crystals, or even monolayer slab, can be obtained through mechanical exfoliation. If this comes true, it might realize the theoretically predicted 2D TIs [4]. We have tested accessibility of 2D atomic crystals of ZrSiSe and ZrSiTe using microexfoliation technique. As shown in

Supplementary Figs. 6a and 6b, in our initial trails, we have already thinned the bulk ZrSiTe single crystals down to 8.6 nm thick flake (equivalent to a stack of 9 slabs), with the lateral dimension reaching > 5μm. Interestingly, our efforts have also produced ZrSiSe thin layers (Supplementary Figs. 6c and 6d) with a thickness of 11nm (13 slabs), despite its more 3D-like structural characteristics. Therefore, ZrSiM compounds hold great potential to fill the gap between the important fundamental physics of topological materials and their practical quantum device applications.

**Methods**

**Single crystal growth**

The ZrSiSe and ZrSiTe single crystals were prepared using a chemical vapor transport method. The stoichiometric mixture of Zr, Si, and Se/Te powder was sealed in a quartz tube with iodine being as transport agent (2 mg/cm$^3$). Plate-like single crystals with metallic luster can be obtained via the vapor transport growth with a temperature gradient from 950 °C to 850 °C. The composition and structure of the single crystals were checked by Energy-dispersive X-ray spectrometer and X-ray diffractometer respectively.

**Magnetization measurements**

The magnetization measurements were performed by a 7T SQUID magnetometer (Quantum Design). Standard copper and quartz holders were used for the out-of-plane and in-plane magnetization measurements, respectively. The sample holder is found to contribute paramagnetic and diamagnetic background for the out-of-plane and in-plane dHvA effects, which, however, does not affect our analyses for the dHvA oscillations after the background subtraction. The angular-dependence of dHvA oscillations were measured using a home-made sample holder.

**Magnetotransport measurements**

The longitudinal and Hall resistivity were measured using standard four and five- probe technique, respectively, in a Physics Properties Measurement System (PPMS). The high field measurements were carried out using the 31 T resistive magnets at National High Magnetic Field Laboratory (NHMFL) in Tallahassee.

**Acknowledgement**

This work was supported by the US Department of Energy under grant DE-SC0014208 (support for personnel, material synthesis and magnetization measurements). The work at the National High Magnetic Field Laboratory is supported by the NSF Cooperative Agreement No. DMR-1157490 and the State of Florida (transport measurements under high magnetic fields). The wok at University of California-Riverside is supported by DOE BES Division under grant no. ER 46940-DE-SC0010597.

**Author contributions**

J. Hu and Z.J.T equally contributed to this work. The single crystals used in this study were synthesized and characterized by J.H, Z.J.T, and Y.L. Z. The magnetization measurements were performed by Z.J.T and J.H. The magnetotransport measurements was carried out by J.Y.L, J.H., Y.S., S. C., D.G. and J.N.L. X.L and J.W exfoliated the thin flakes. J.H., Z.J.T, and Z.Q.M conducted data analyses. J.H and Z.Q.M wrote the manuscript. This project was supervised by Z.Q.M.

**Competing financial interests:** The authors declare no competing financial interests.

# Figure captions

**Figure 1 | Crystal structure and dHvA oscillations of ZrSiSe and ZrSiTe. a,** Crystal structure of ZrSiM (M=S, Se, Te). **b,** Single crystal X-ray diffraction spectra of ZrSiM (M=S, Se, Te). From the (00L) diffraction peaks, the *c*-axis lattice parameters (*i.e.* the slab thickness) are obtained. **c,** Optical images of single crystals of ZrSiSe and ZrSiTe. **d** and **e,** Isothermal out-of-plane (*B*//*c*, panel **d**) and in-plane (*B*//*ab*, panel **e**) magnetization (*M*) for ZrSiSe at various temperatures from *T*=1.8K to 20K. Strong dHvA oscillations are clearly seen. The same temperature - color correspondence is used for both panels. **f,** Isothermal out-of-plane (*B*//*c*) magnetization (*M*) of ZrSiTe at various temperatures from *T*=1.8K to 18K.

**Figure 2 | Analyses for the dHvA oscillations for ZrSiSe. a** and **d,** The oscillatory component of the (**a**) out-of-plane (*B*//*c*) and (**d**) in-plane (*B*//*ab*) magnetization Δ*M* for ZrSiSe, obtained by subtracting the paramagnetic background of the magnetization. **b** and **e,** the fits of the FFT amplitudes (see their insets) to the temperature damping term $R_T$ of the LK formula (see text). Effective masses associated with various frequencies for out-of-plane and in-plane oscillations can be obtained from the fits, as listed in Table 1. The insets show the FFT for Δ*M* shown in **a** and **d**, respectively. **c,** The fit (red line) of the oscillation pattern (blue points) at *T*=1.8K to the LK formula (Eq. 1, see text). From this fit, the Dingle temperature and Berry phase can be obtained. To better illustrate the fitting quality, the zoomed-in data and fitting is shown in the inset. **f,** the low-(upper panel) and high-(lower panel) oscillation components of the in-plane dHvA oscillations. The red solid line, which overlaps with the experimental data points, is the fit of the oscillation pattern to the multiband LK formula (see text). From this fit, the Dingle temperature and Berry phase can be obtained, as listed in Table 1.

**Figure 3 | Analyses for the dHvA oscillations for ZrSiTe. a,** The oscillatory component of the out-of-plane ($B//c$) magnetization $\Delta M$ for ZrSiTe, obtained by subtracting the paramagnetic background. **b,** The fits of the FFT amplitudes (see their insets) to the temperature damping term $R_T$ of the LK formula (see text). Effective mass of $0.093m_0$ and $0.091m_0$ can be obtained for the two frequencies 102T and 154T, respectively. The inset shows the FFT spectrum of $\Delta M$ shown in **a**. **c,** The fit (red line) of the oscillation pattern (blue points) of ZrSiTe at $T=1.8K$ to the LK formula (Eq. 1, see text).

**Figure 4 | Fermi surface morphology of ZrSiSe and ZrSiTe. a** and **c,** dHvA oscillations of (**c**) ZrSiSe and (**b**) ZrSiTe at $T=1.8K$ for different magnetic field orientations. The field is rotated from the out-of-plane direction ($B//c$, defined as $\theta=0°$) to the in-plane direction ($B//ab$, defined as $\theta=90°$), as shown in the inset in **b**. For ZrSiTe, dHvA oscillations are too weak to be resolved above $\theta=45°$. The data collected under different field orientations have been shifted for clarity. **b** and **d,** the angular dependences of the oscillation frequencies for (**b**) ZrSiSe and (**d**) ZrSiTe, obtained from the FFT spectra (Supplementary Fig. 3) of the dHvA and SdH oscillations. Error bars are defined as the half-width at the half-height of FFT peak. The red lines are fits to $F = F_{3D}+F_{2D}/\cos\theta$. **e,** The normalized in-plane magnetoresistance $\Delta\rho_{xx}/\rho_0$, defined as $[\rho_{xx}(B)-\rho_{xx}(B=0)]/\rho_{xx}(B=0)$, for ZrSiSe at $T = 2K$, measured up to 31T under different magnetic field orientations. SdH oscillations can be seen. **f,** Magnetic field dependence of Hall resistivity $\rho_{xy}$ for ZrSiSe. The non-linear field dependence at low fields implies multiband feature of ZrSiSe. SdH oscillations are also seen at low temperatures (below 20K). **f,** Temperature of Hall coefficient $R_H$, extracted from the slope of $\rho_{xy}(B)$ at 9T.

**Table 1 Quantum oscillation properties for ZrSiSe and ZrSiTe.** The oscillation frequency $F$, Dingle temperature $T_D$, effective mass $m^*/m_0$, quantum relaxation time $\tau_q$ [$= \hbar/(2\pi k_B T_D)$], quantum mobility $\mu_q$ ($= e\tau/m^*$), and Berry phase $\phi_B$ of different Dirac bands probed by dHvA oscillations.

| | | $F$ (T) | $T_D$ (K) | $m^*/m_0$ | $\tau$ (ps) | $\mu$ (cm$^2$/Vs) | $\phi_B$ $\delta = -1/8$ | $\delta = 0$ | $\delta = 1/8$ |
|---|---|---|---|---|---|---|---|---|---|
| ZrSiSe | $B//c$ | 210 | 5.6 | 0.082 | 0.22 | 4605 | 0.31π | | -0.19π |
| | $B//ab$ | 19.2 | 5.8 | 0.039 | 0.21 | 9461 | 1.17π | 0.92π | 0.67π |
| | | 22.8 | 8.2 | 0.042 | 0.15 | 6214 | 1.69π | 1.44π | 1.19π |
| | | 24 | 11.1 | 0.037 | 0.11 | 5211 | 1.65π | 1.4π | 1.15π |
| | | 126.9 | 10.8 | 0.059 | 0.11 | 3359 | 1.15π | 0.9π | 0.65π |
| | | 132.7 | 14.6 | 0.057 | 0.08 | 2572 | 1.75π | 1.5π | 1.25π |
| | | 142 | 5 | 0.078 | 0.244 | 5488 | 1.62π | 1.37π | 1.12π |
| ZrSiTe | $B//c$ | 102 | 14 | 0.093 | 0.086 | 1625 | 0.75π | 0.5π | 0.25π |
| | | 154 | 25 | 0.091 | 0.049 | 940 | 1.55π | 1.3π | 1.05π |

Figure 1

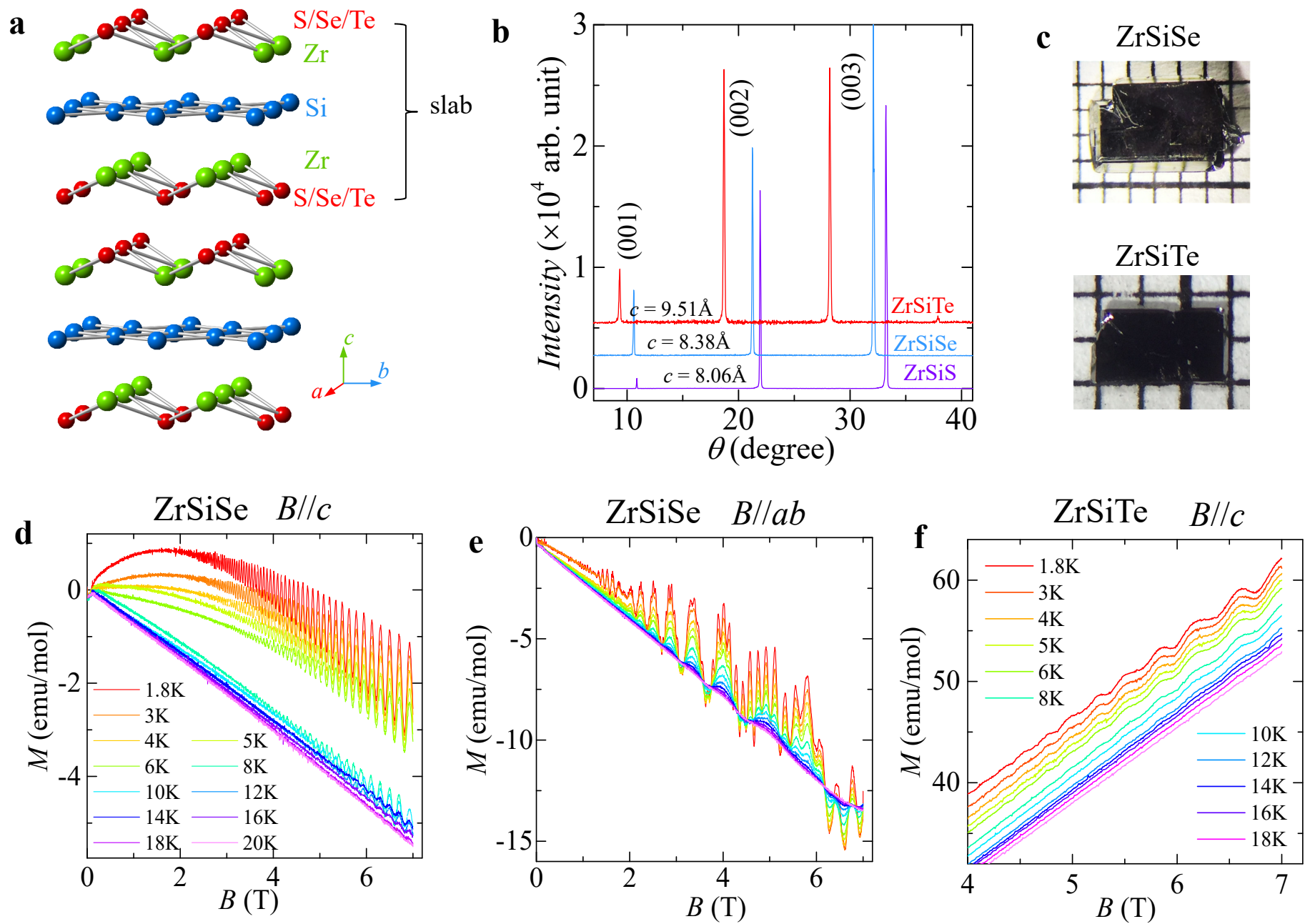

Figure 2

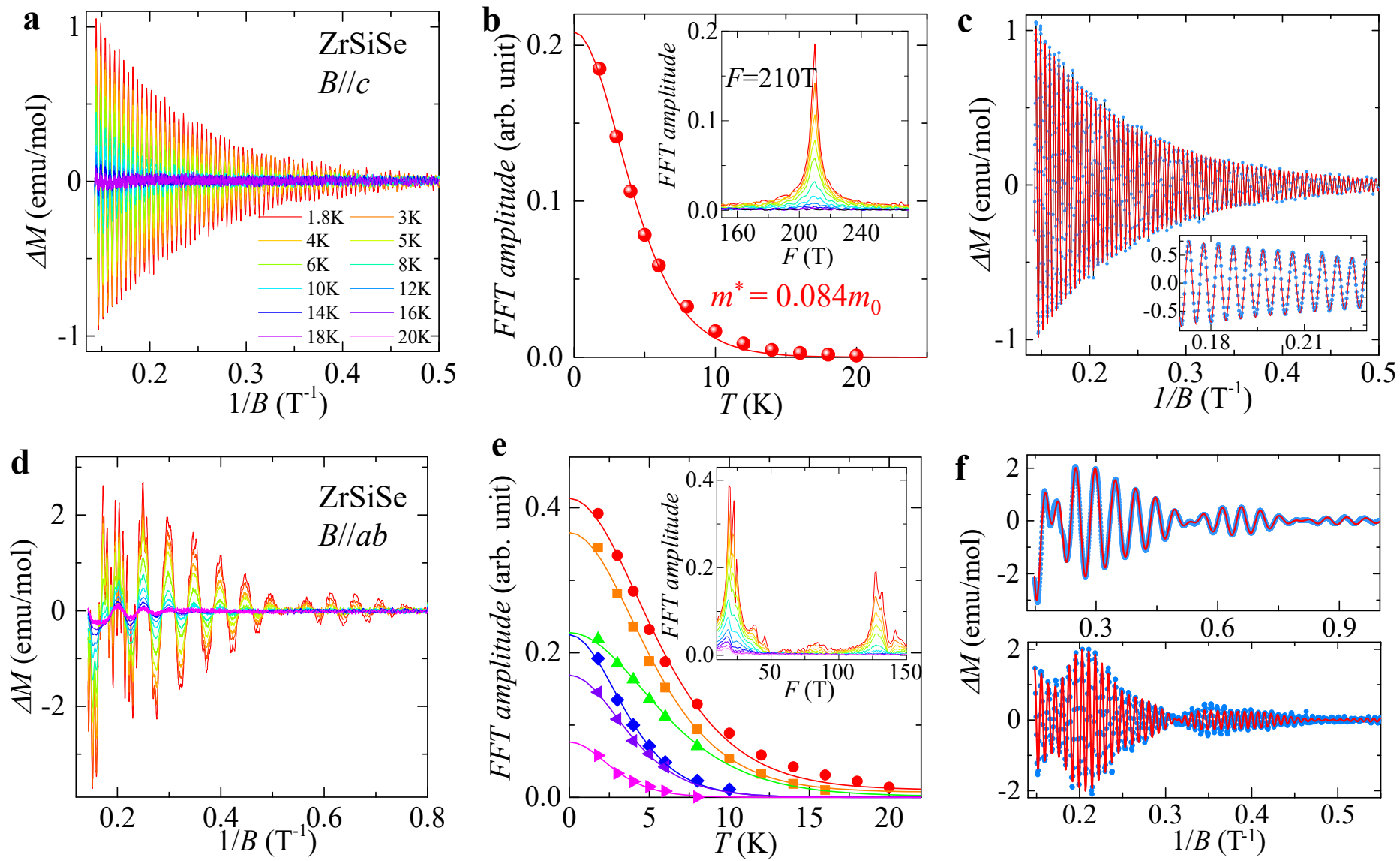

Figure 3

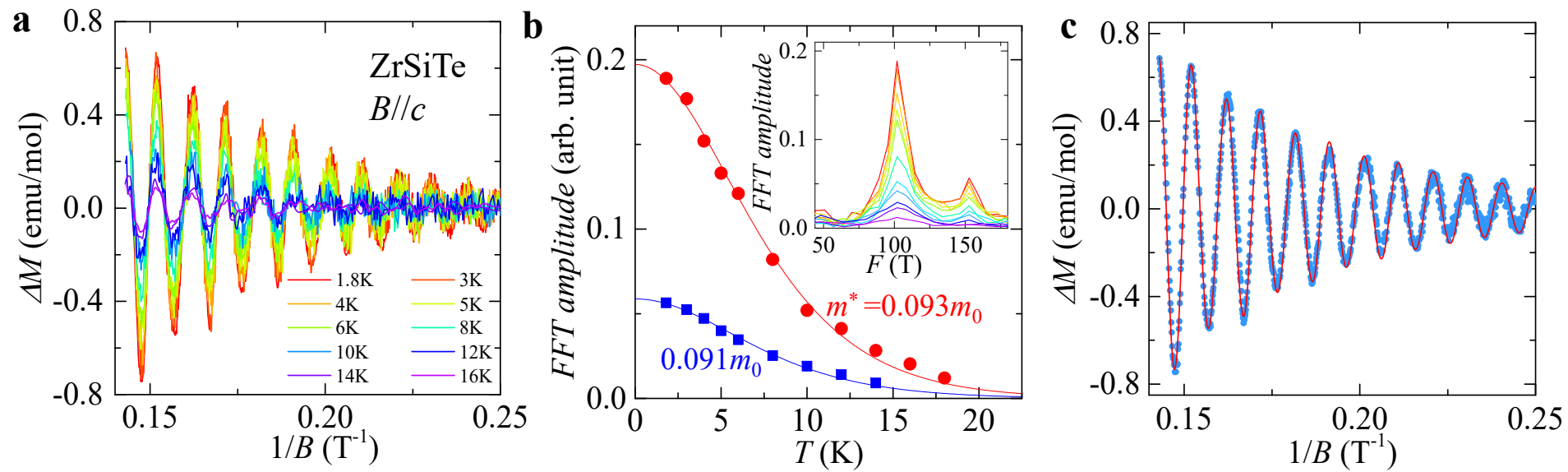

Figure 4

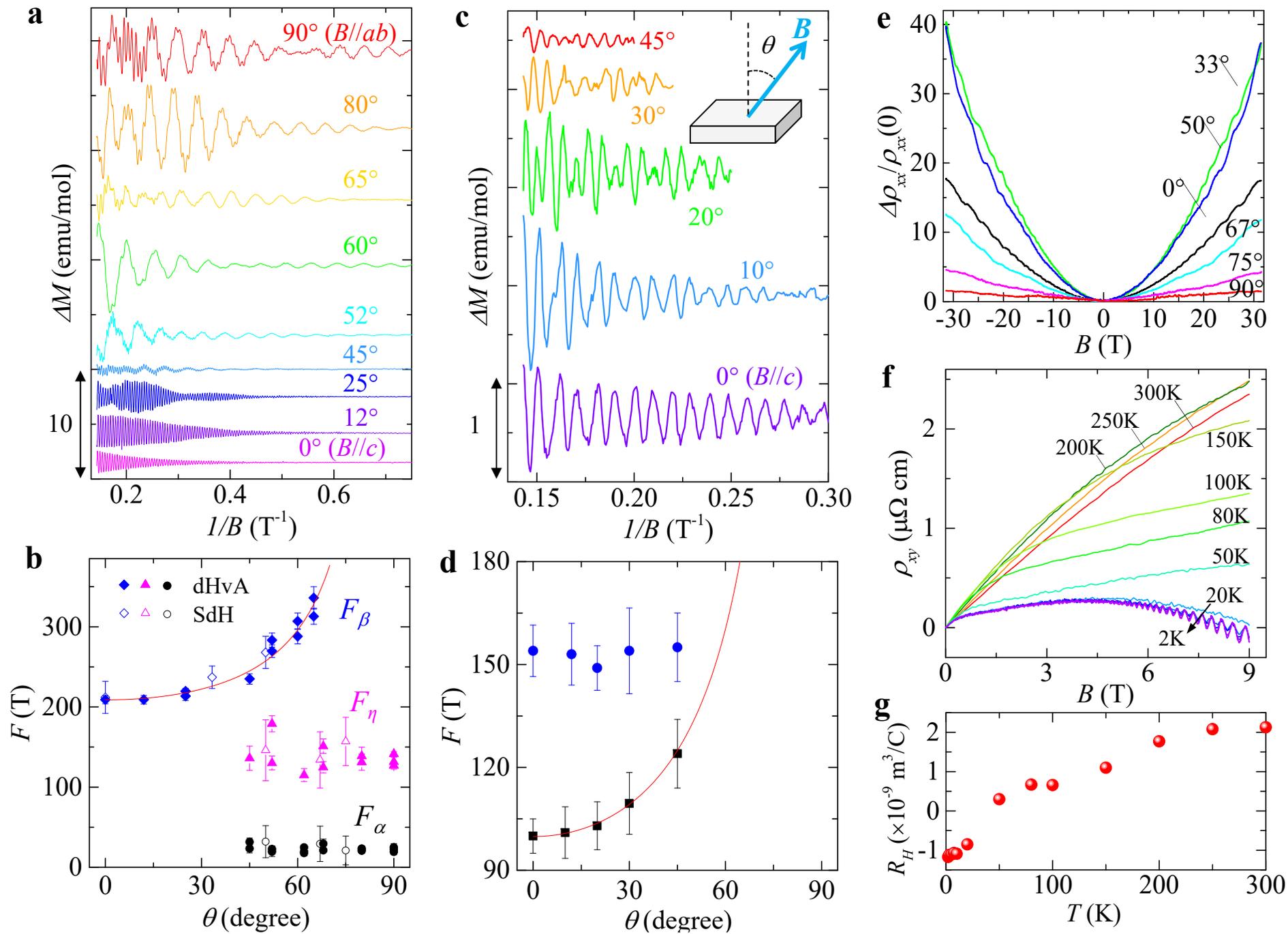

# Supplementary Information for

# Topological nodal-line fermions in ZrSiSe and ZrSiTe

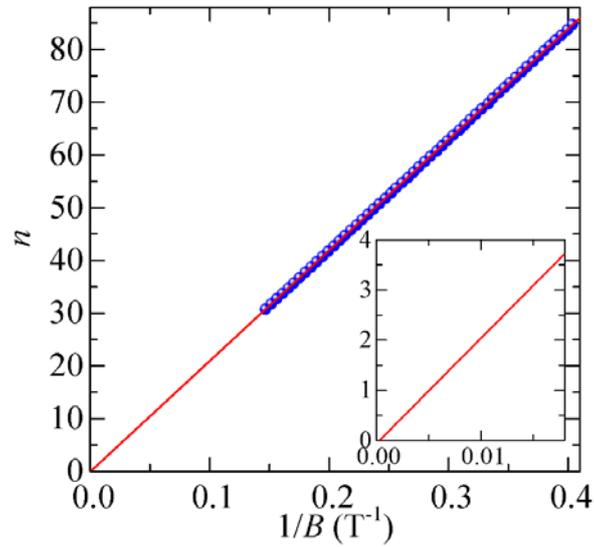

**Supplementary Figure 1** | Landau Level (LL) index fan diagram for the out-of-plane (*B//c*) dHvA oscillations. The landau indices of the oscillation minimum of $\Delta M$ is assigned to be *n*-1/4. The oscillation frequency obtained from linear fit is 209.96T, in good agreement with the FFT result. The intercept derived from the fit is 0.06, corresponding to a Berry phase of $2\pi(0.06+\delta)$, i.e., non-trivial for 3D Fermi surface.

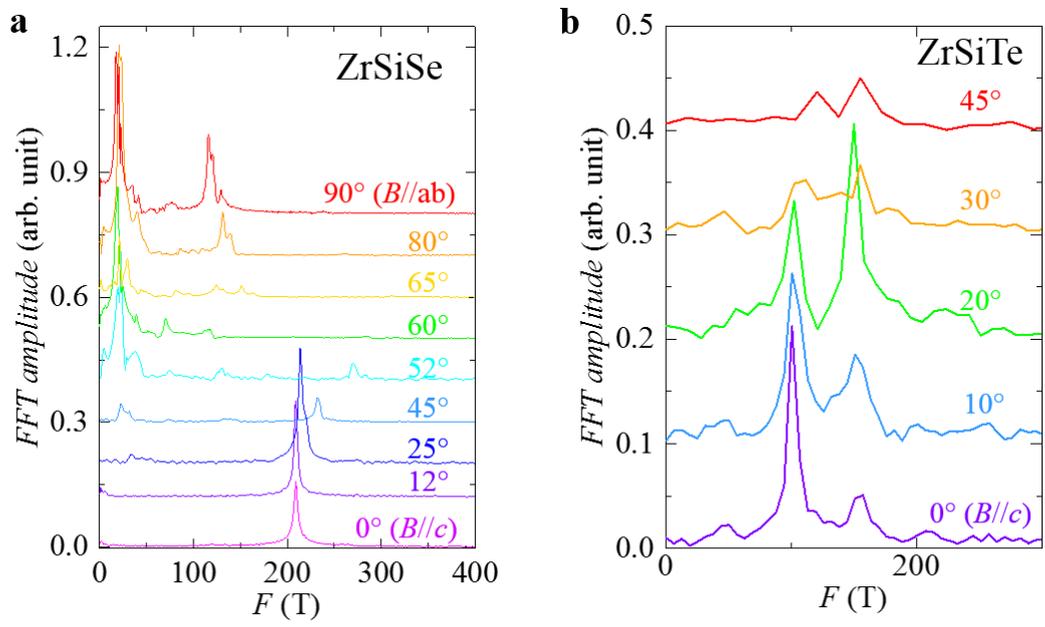

**Supplementary Figure 2 |** FFT of the out-of-plane dHvA oscillations under different magnetic field orientations at $T$=1.8K for (a) ZrSiSe and (b) ZrSiTe.

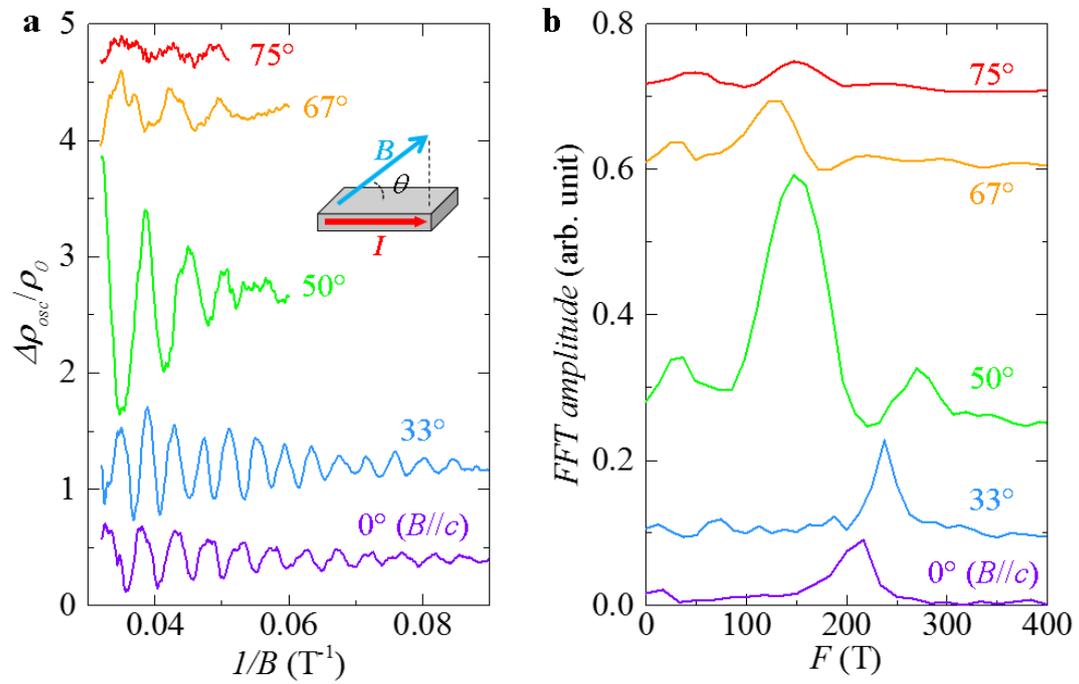

**Supplementary Figure 3 | a,** Oscillatory component of the SdH oscillations under various magnetic field orientation at $T$=2K, obtained after removing the background of the in-plane magnetoresistance shown in Fig. 4e. The inset shows the measurement setup. **b,** FFT of the SdH oscillatory component.

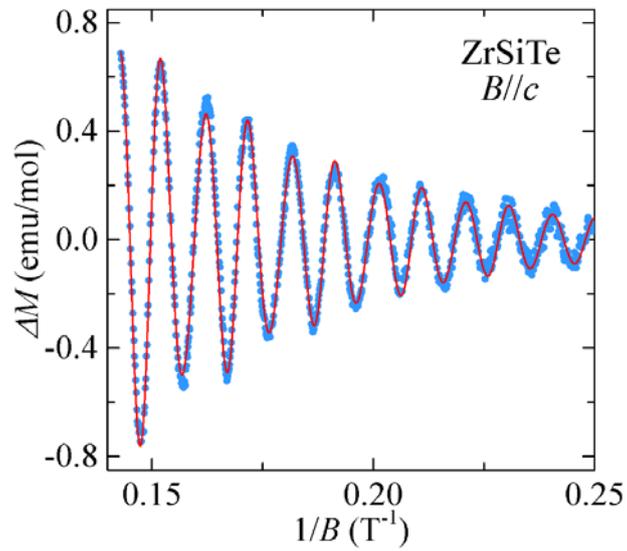

**Supplementary Figure 4 | Fit of the out-of-plane (*B//c*) dHvA oscillation pattern of ZrSiTe to the 2D LK formula.** The obtained phase factor $\gamma - \delta$ is almost the same between the 2D ($\delta=0$) and 3D ($\delta=1/8$) LK formula fits. However, the Dingle temperature are slightly different, 15.3K and 24.1K for the 102T and 154T components, respectively.

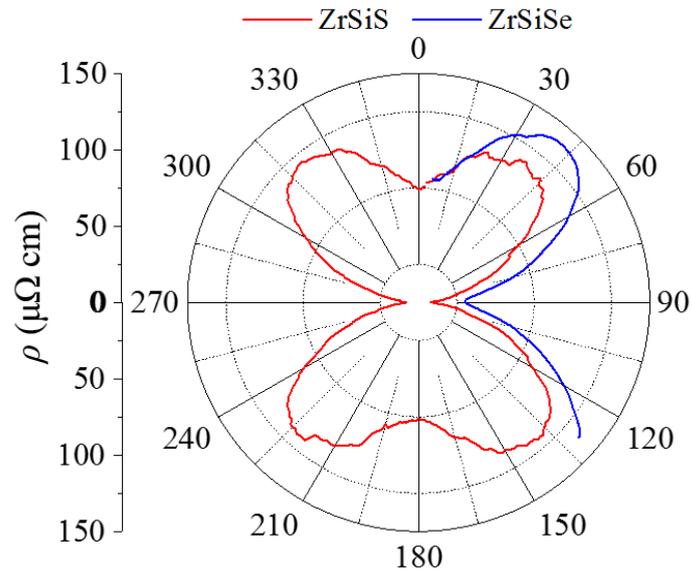

**Supplementary Figure 5 | Butterfly shape anisotropic in-plane magnetoresistance of ZrSiM (M=S, Se).** The measurement setup is shown in the inset of the Supplementary Figure 3a. The angular-dependence of the in-plane magnetoresistance at 2K and 9T exhibits a butterfly shape in the polar plot for ZrSiS, which is also seen in other reports. The magnetoresistance of ZrSiSe, measured at 1.9K and 10T, exhibits a similar anisotropy.

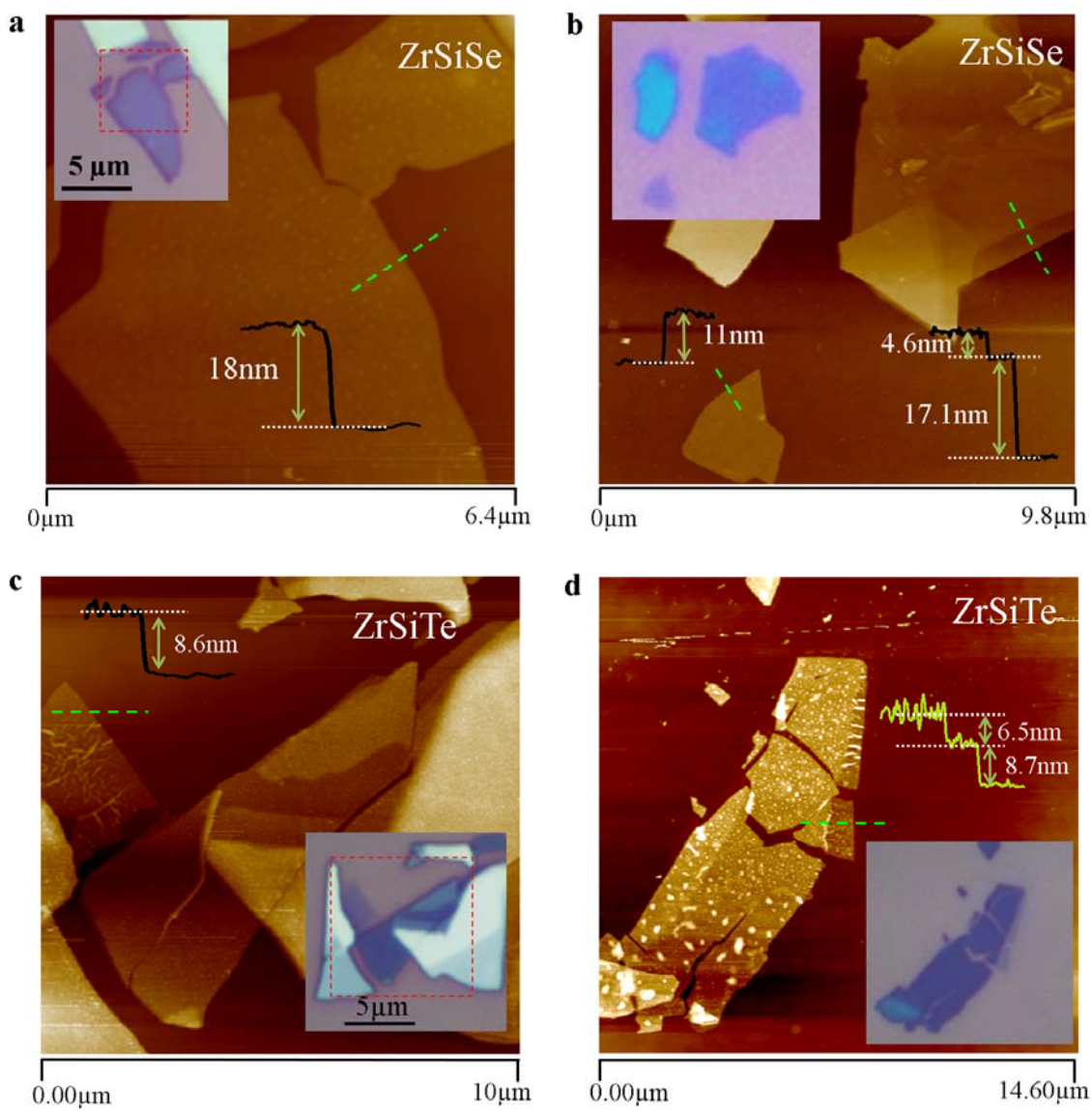

**Supplementary Figure 6 | ZrSiSe and ZrSiTe thin flakes obtained using micro-exfoliation method. a** and **b,** Atomic force microscope (AFM) images of exfoliated (**a** and **b**) ZrSiSe and (**c** and **d**) ZrSiTe thin flakes. The corresponding optical images are shown in the (**a** and **b**) upper left and (**c** and **d**) lower right insets. The flake thickness has been determined by the height profile from the AFM scan along the green dashed lines.